# Thrombogenic Risk Assessment of Transcatheter Prosthetic Heart Valves Using a Fluid-Structure Interaction Approach

Authors: Kyle Baylous[1], Brandon Kovarovic[1], Salwa Anam[1], Ryan Helbock[1], Marvin Slepian[2], Danny Bluestein[1]*

[1]Department of Biomedical Engineering, Stony Brook University, Stony Brook, NY, USA 11794

[2]Department of Medicine and Biomedical Engineering Sarver Heart Center, University of Arizona, Tucson, AZ 85721

\* Corresponding Author Contact Information

Dr. Danny Bluestein

Department of Biomedical Engineering, Stony Brook University

T8-050 Health Sciences Center

Stony Brook, NY 11794-8084, USA

Tel: +1 (631) 444-2156

Email: danny.bluestein@stonybrook.edu

**ABSTRACT**

***Background and Objective:*** Prosthetic heart valve interventions such as TAVR have surged over the past decade, but the associated complication of long-term, life-threatening thrombotic events continues to undermine patient outcomes. Thus, improving thrombogenic risk analysis of TAVR devices is crucial. In vitro studies for thrombogenicity are typically difficult to perform. However, revised ISO testing standards include computational testing for thrombogenic risk assessment of cardiovascular implants. We present a fluid-structure interaction (FSI) approach for assessing thrombogenic risk of prosthetic heart valves.

***Methods:*** An FSI framework was implemented via the incompressible computational fluid dynamics multi-physics solver of the Ansys LS-DYNA software. The numerical modeling approach for flow analysis was validated by comparing the derived flow rate of the 29-mm CoreValve device from benchtop testing and orifice areas of commercial TAVR valves in the literature to in silico results. Thrombogenic risk was analyzed by computing stress accumulation (SA) on virtual platelets seeded in the flow fields via Ansys EnSight. The integrated FSI-thrombogenicity methodology was subsequently employed to examine hemodynamics and thrombogenic risk of TAVR devices with two approaches: 1) engineering optimization and 2) clinical assessment.

***Results:*** The simulated effective orifice areas of the commercial devices were in the range reported in the literature. The flow rates from the in vitro flow testing matched well with the in silico results. The approach was used to analyze the effect of various TAVR leaflet designs on hemodynamics. Platelets experienced different magnitudes of SA along their trajectories as they flowed past each design. Post-TAVR deployment hemodynamics in patient-specific bicuspid aortic valve anatomies revealed varying degrees of thrombogenic risk for these patients, despite being clinically defined as "mild" paravalvular leak.

***Conclusions:*** Our methodology can be used to improve the thromboresistance of prosthetic valves from the initial design stage to the clinic. It allows for unparalleled optimization of devices, uncovering key TAVR leaflet design parameters that can be used to mitigate thrombogenic risk, in addition to patient-specific modeling to evaluate device performance. This work demonstrates the utility of advanced in silico analysis of TAVR devices that can be utilized for thrombogenic risk assessment of other blood recirculating devices.

**Abbreviations:**

| | |
|---|---|
| AS: Aortic stenosis | BAV: Bicuspid Aortic Valve |
| EOA: Effective Orifice Area | GOA: Geometric Orifice Area |
| FEM: Finite Element Modeling | ALE: Arbitrary Lagrangian-Eulerian |
| FSI: Fluid-Structure Interaction | FDA: Food and Drug Administration |
| PDF: Probability Density Function | PVL: Paravalvular Leak |
| SAVR: Surgical Aortic Valve Replacement | TAVR: Transcatheter Aortic Valve Replacement |
| SA: Stress Accumulation | ISO: International Organization for Standardization |

## 1. INTRODUCTION

Transcatheter aortic valve replacement (TAVR) is a well-established, minimally invasive treatment for aortic valve disease in patients with severe aortic stenosis (AS) [1]. Despite the efficacy of TAVR, concerns about device-related thrombosis persist [2-5]. Currently, TAVR is associated with several clinical complications, including long-term, life-threatening thrombotic events from exposure of platelets to elevated cyclic stresses. Flow-induced thrombogenicity increases the risk of platelet activation, and microthromboemboli formation can lead to thrombosis and elevated stroke risk, in addition to device failure [6-9].

TAVR devices undergo testing to ensure safety via the ISO 5840 standard [10], including thrombogenic potential assessment with no established testing criteria. The manufacturers establish and justify their approach and techniques used for assessing their devices thrombogenicity. This highlights the challenges associated with conducting in vitro thrombogenicity studies: replicating physiological conditions, limited blood resources, difficulties in evaluating thrombus formation, and the inability to predict long-term thrombogenicity of biomaterials [11]. Current approaches to assess thrombogenic risk predominantly rely on clinical parameters and imaging modalities with their inherent limitations [12].

In silico modeling has emerged as a promising method for evaluating thrombogenic potential. The updated ISO testing standards now incorporate computational testing for assessing the thrombogenic risk of cardiovascular implants, with in vitro and in silico approaches currently serving as complementary methods. [13]. Advanced in silico approaches can be used to elucidate intricate flow patterns and shear stresses within the vicinity of implanted devices with high granularity. Such approaches have been developed and utilized in previous work to examine performance of prosthetic devices. Bluestein et al. studied shear mediated platelet activation (SMPA) of implantable left ventricular assist devices (LVAD) and prosthetic heart valves [14, 15]. It was previously applied to a trileaflet polymeric valve for surgical aortic valve replacement (PolyNova Cardiovascular Inc., Stony Brook, NY) using FSI [16]. This in silico methodology for predicting the thrombogenic potential of blood recirculating devices, named Device Thrombogenicity Emulation (DTE), was validated in LVAD both in vitro and in vivo [17]. Computational fluid dynamics has been used for thrombogenic risk assessment of prosthetic heart valves [18, 19]. Such simulation approaches are essential for assessing TAVR devices thrombogenicity, given the propensity for thrombus formation and the associated risk of stroke that may hamper their long-term success. This will be exacerbated as TAVR expands to younger patient cohorts [20]. It jeopardizes patient outcomes and underscores the pressing need to capture the intricate interplay of hemodynamic forces, engineered design of prosthetic devices, and patient-specific factors that contribute to thrombogenicity and complicate risk assessment and lifetime management.

In this study we assess TAVR device performance and thrombogenic risk using an advanced FSI framework. We first apply it to analyze effects of TAVR leaflet design on hemodynamic performance and thrombogenic risk using a parametric leaflet design that permits optimization of polymeric valves (Fig. 1ab) [21]. We then use the FSI approach to analyze the risk of post-TAVR thrombogenicity in patient-specific BAV models, as this patient cohort has significant occurrence of stroke (Fig. 1c) [22]. Our approach enables thrombogenicity evaluation and can enhance TAVR devices by assessing their hemodynamic performance on a patient-specific basis and modifying their designs to optimize thromboresistance.

## 2. METHODS

### *2.1 Fluid-Structure Interaction (FSI) Model Setup*

The FSI analysis was performed with a strong coupling, implicit technique with the program LS-DYNA (Ansys, Canonsburg, PA). The analysis captures valve dynamics in both idealized and patient-specific geometries– for TAVR leaflet design and patient-specific comparisons, respectively (Fig. 1). Anonymized pre-TAVR cardiac CT scans of six type I BAV patients who underwent TAVR were obtained from Rabin Medical Center (Stony Brook IRB protocol 522113) and received 29-mm self-expandable TAVR devices (Medtronic plc, Minneapolis, MN, USA). Patient models were created as described in our previous work [18]. TAVR deployments were simulated to replicate the patient procedures using ABAQUS (Dassault Systèmes, SIMULIA Corp.). The post-deployment geometries were then taken into Ansys Fluent for mesh editing and loaded into LS-PrePost for model and keyword configuration before FSI simulation in LS-DYNA (Fig. 1).

The FEM-ALE (Finite Element Method - Arbitrary Lagrangian–Eulerian) approach is utilized here, as described in the literature [23, 24]. In all simulations, the fluid domain boundary was assumed rigid. Inflow dimensions were 1.5 times the left ventricular outflow tract diameter while the outflow dimensions were 3.5 times aortic annulus dimensions [25]. A physiological pressure gradient generated in MATLAB (MathWorks, Natick, MA) was applied at the inlet, with zero pressure at the outlet. No-slip conditions were imposed at the walls and all leaflet surfaces. The blood flow was simulated as Newtonian fluid with a density of 1,060 kg/m$^3$ and viscosity of 0.0035 Pa-s. The fluid domain volume elements were remeshed continuously to handle leaflet deflection during the cardiac cycle. The timestep was allowed to vary between values of 1E-4 and 2E-4 seconds, respectively, consistent with our previous studies and those reported by others [24, 26]. For leaflet modeling, fully integrated shell elements were utilized and embedded in the fluid domain. The nodes defining the leaflet attachment region to the stent were fixed.

The Mooney-Rivlin and Ogden hyperelastic material models were implemented to define leaflet material properties of the polymeric valve designs, modeled by using xSIBS material properties (PolyNova Cardiovascular Inc., Stony Brook, NY) [21], and commercial TAVR devices modeled using pericardial bioprosthetic tissue properties [27].

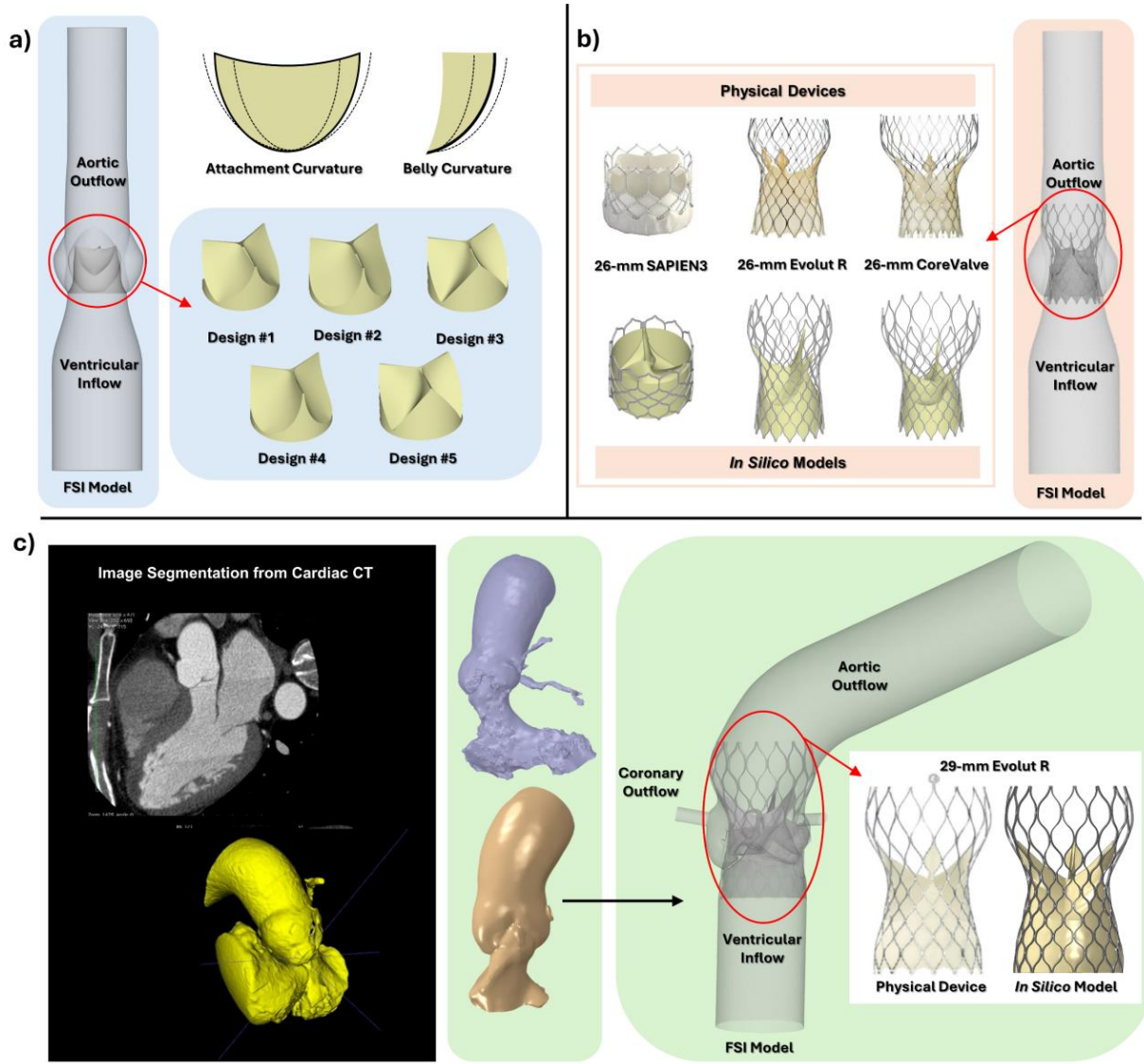

**Figure 1.** a) Illustration of the FSI model setup and polymeric leaflet designs; b) Physical and virtual commercial TAVR devices (26-mm SAPIEN3, Evolut R and CoreValve) used for the FSI simulations to compare to in-house leaflet designs; c) BAV patient anatomy segmentation and model construction for the patient-specific FSI simulations for post-TAVR hemodynamics of the 29-mm Evolut R device.

*2.2 FSI Model Validation*

To validate the FSI model setup and approach used, 1) the derived flow rate and high speed imaging snapshots (1057 FPS, Chronos 1.4, Kron Technologies Inc, Burnaby, BC) from benchtop hydrodynamic testing of a 29-mm CoreValve device were compared to in silico results and 2) simulated effective orifice area (EOA) values for three commercial TAVR devices (26-mm SAPIEN3, Evolut R and CoreValve) were compared to data reported in the literature. The hydrodynamic testing of a 29-mm CoreValve was conducted using a pulse duplicator system (Fig. 2a) (Left Heart Simulator, ViVitro Labs, Victoria, BC). The device was deployed in a silicone hydrodynamic test fixture following ISO 5840-3 standards, simulating an idealized deployment model once placed in a testing chamber with aortic root sinus geometry (Fig. 2bcd). The test fluid used was the blood analog glycerol, and in vitro testing was carried out with the following parameters: a heart rate of 70 BPM, aortic pressure of 120/80 mmHg, cardiac output of 3 L/min and a temperature of 37°C. Valve dynamics were captured with a high-speed camera orientated to capture the outlet. The hydrodynamic testing setup was replicated virtually (Fig. 2ef). The recorded pressure gradient from the hydrodynamic testing was used as a boundary condition for validating the FSI. In addition to the benchtop testing, the simulated EOA for the 26-mm devices were validated with published data [28, 29] after carrying out idealized TAVR deployments in ABAQUS and performing FSI with the post-deployment models.

*2.3 Implementation of Thrombogenic Risk Assessment*

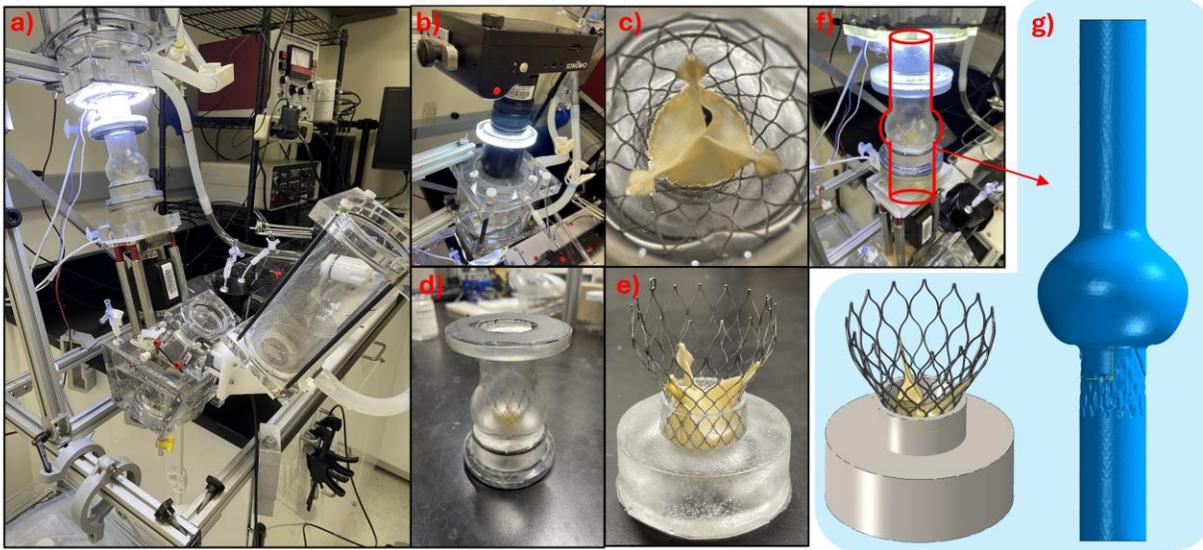

**Figure 2.** a) Pulse duplicator system setup (ViVitro Labs Inc.); b) High speed camera positioning for capturing valve dynamics; c) Overhead snapshot of deployed 29-mm CoreValve in hydrodynamic testing fixture seated within chamber featuring sinus geometry; d) Side view of deployed 29-mm CoreValve; e) Snapshot of device placement within the silicone hydrodynamic testing fixture; f) Simplified illustration of hydrodynamic testing geometry used for virtual model creation overlaid on pulse duplicator system; g) simulation geometry and virtual device deployment mimicking benchtop testing.

Thrombogenic potential was assessed via a Lagrangian particle tracking approach where virtual platelet trajectories were computed via pathlines and assumed to be massless due to the scale of the fluid domain with respect to an individual platelet. This approach involves seeding the virtual platelets at a predefined cross section in the fluid domain during post-processing in Ansys EnSight, using the flow results of the FSI simulations to compute Lagrangian trajectories and track the velocity and stress data for each platelet (Fig. 3). Approximately 20,000 platelets were uniformly seeded in each simulation. Seeding commenced at the onset of the first cycle, recurring every 5 ms for the entire positive systolic pressure gradient. Scalar stress values were computed using a MATLAB code (Eq. 1) after processing stress component data in EnSight using a python script.

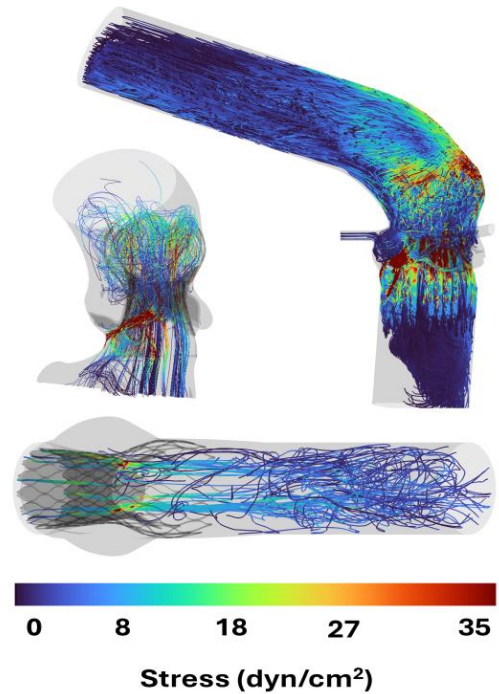

**Figure 3.** Example images showing virtual platelet tracking with contour plot for stress overlaid on the trajectories.

$$\sigma = \frac{1}{\sqrt{3}}\sqrt{\sigma_{11}^2 + \sigma_{22}^2 + \sigma_{33}^2 - \sigma_{11}\sigma_{22} - \sigma_{11}\sigma_{33} - \sigma_{22}\sigma_{33} + 3(\tau_{12}^2 + \tau_{23}^2 + \tau_{13}^2)} \qquad (1)$$

The scalar stress value expresses the contribution of shear and principal stresses to represent the total instantaneous stress loading acting on the platelets (von-Mises stress), utilized in calculating stress accumulation (SA) on each platelet (Eq. 2) [15, 30, 31].

$$SA = \sigma \cdot t_{exp} = \int_{t_0}^{t_{exp}} \sigma(t)dt \approx \sum_{i=1}^{N} \sigma_i \cdot \Delta t \qquad (2)$$

All particle SA along trajectories were aggregated into a probability density function (PDF), indicating the likelihood of platelets reaching their activation threshold [15, 16]. Each generated PDF curve, whether it be for a different TAVR leaflet design or BAV patient who underwent TAVR, serves as the thrombogenic footprint for each respective scenario, enabling comparative analysis of the thrombogenic potential by examining PDFs. This approach is robust as the activation percentage is independent of the number of seeded particles and spatiotemporal variations, since the PDF approach employs bootstrapping statistics.

*2.4 Thrombogenic Risk Analysis of TAVR Leaflet Designs*

Commercial TAVR models were generated as described in our previous work [32] and multiple polymeric valve designs were created by varying leaflet attachment and belly curvature, allowing for comparative analysis of the thrombogenic footprint of the polymeric devices with respect to commercial devices, which the FSI model was validated against. With these simulations, idealized streamlined TAVR deployments were generated and FSI analysis was performed for two cardiac cycles. The initial fluid mesh comprised 1.5-2 M tetrahedral volume elements, which decreased to 1-1.2 M during successive remeshing steps. This mesh density range established mesh independence, being four times finer than meshes used in previous FSI work [16, 24]. Virtual platelets were seeded during the systolic, forward flow phase of the cardiac cycle (Fig. 4) during the second beat to elucidate how TAVR leaflet geometry affects valve dynamics and thrombogenic risk.

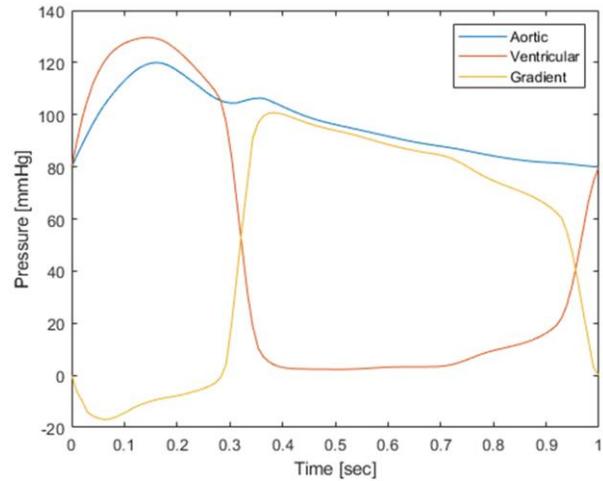

**Figure 4.** Plot showing generation of physiological pressure gradient curve in MATLAB by computing aortic and ventricular waveforms and then taking the difference between the curves. This was used as a boundary condition for all FSI simulations except the benchtop validation studies.

*2.5 Thrombogenic Risk Analysis of BAV Patients Post-TAVR*

The post-TAVR patient geometries were imported into Ansys Fluent and processed to produce an FSI-ready fluid domain, including PVL gaps and coronaries (Fig. 5). The post-TAVR anatomies were then configured for FSI simulations as described previously, with the addition of coronary artery flow involving prescribed velocity conditions derived from the literature using MATLAB [33, 34]. For the BAV patient simulations, the initial fluid mesh consisted of 5-7 M tetrahedral volume elements, which decreased to 2.5-3 M during successive remeshing steps. This mesh density range established mesh independence as described previously. FSI simulations were performed for two cardiac cycles. Virtual platelets were tracked as they flowed past the prosthetic valve in addition to their recirculation through PVL channels. Hellums criteria was used as the platelet activation threshold: 35 dyn-s/cm$^2$ [18].

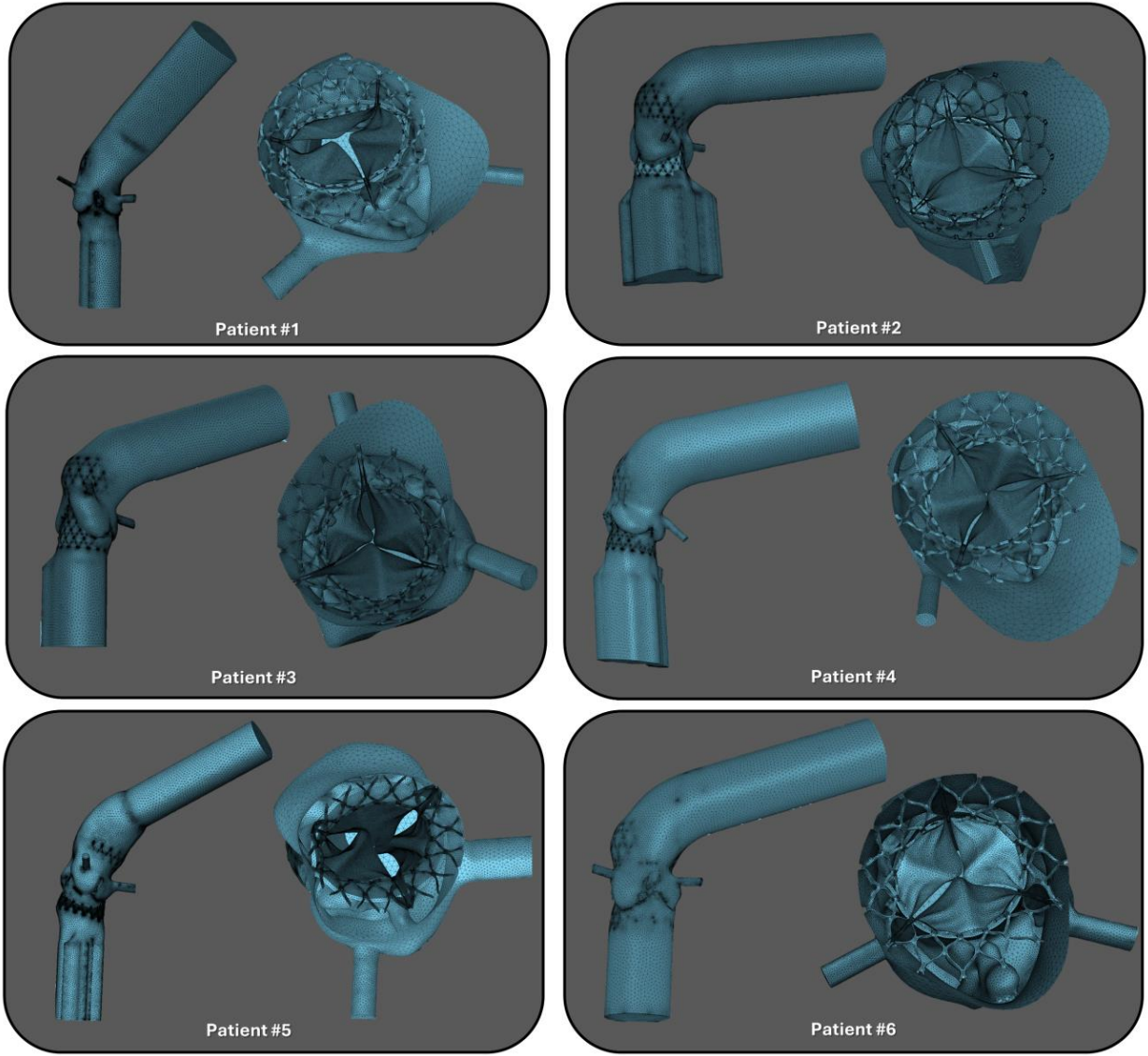

**Figure 5.** The six generated post-TAVR deployment patient-specific anatomy models utilized for the FSI simulations, the left view in each inset image shows the FSI-ready anatomy while the right view shows a cross-sectional view looking down towards the valve along the z-axis.

## 3. RESULTS

### *3.1 Validation Studies*

The in vitro measured pressure gradient was used as the boundary condition for the FSI simulations (Fig. 6a). The flow rate curve recorded during the in vitro experiments shows an appropriate profile for heart valve function (Figure 6b). The in silico results replicating the hydrodynamic testing conditions closely resemble the experimental data (Figure 7). The flow behavior during systole exhibits significant similarity, while during diastole the flow rate approaches zero (Figure 7b). Importantly, the valve dynamics are reproduced in the model, where snapshot images were compared side-by-side for both models (Fig. 7a). The successive configurations of the valve throughout the cycle exhibit excellent agreement. The FSI results for the 26-mm commercial devices exhibit the expected valve leaflets motion, with flow velocities reaching around 2.5 m/s (Fig. 8). Furthermore, the EOAs calculated for these devices agree with values reported in the literature (Table 1) [28, 29].

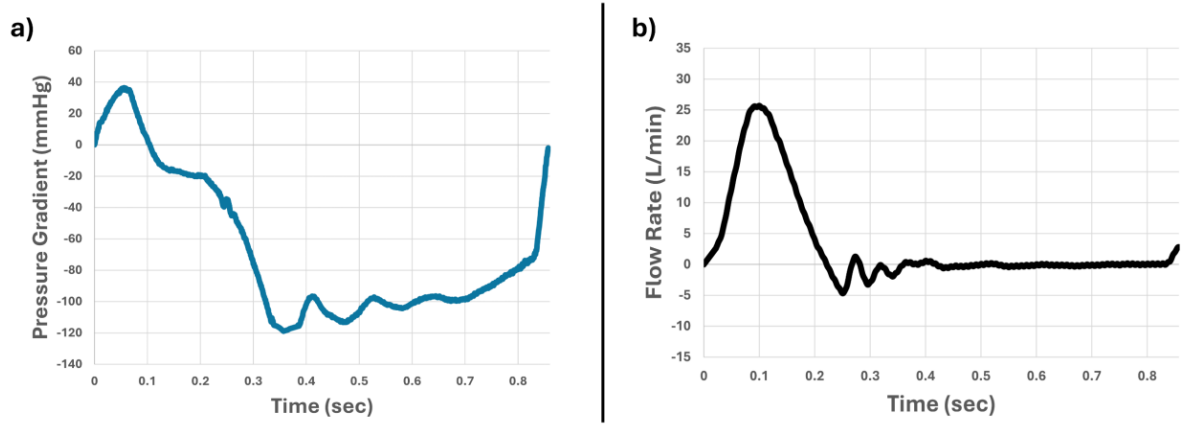

**Figure 6.** a) Derived pressure gradient from in vitro experiments, used as boundary condition for validating the FSI simulation approach; b) Derived flow rate from in vitro experiments, used for validating FSI results.

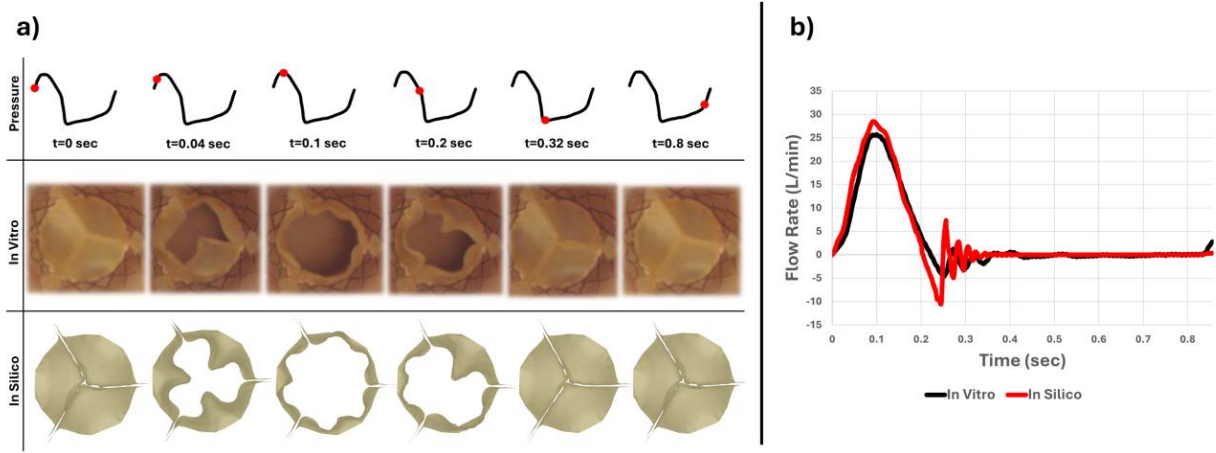

**Figure 7.** a) In silico valve dynamics agree with experimental results, where the dots on the pressure curves indicate to which instant of the cycle the results refer; b) In silico flow rate data agrees with experimental results.

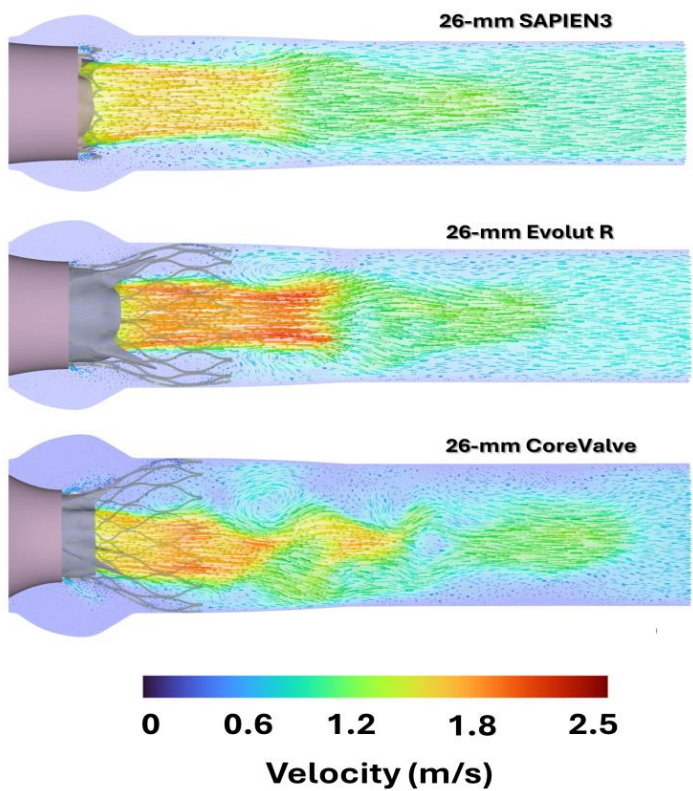

**Figure 8.** FSI results for the 26-mm SAPIEN3, Evolut R and CoreValve devices. Fluid velocity contours and vectors are plotted.

| Device | Simulated EOA | Accepted EOA |
|---|---|---|
| 26-mm SAPIEN3 | 2.06 cm² | 1.74-2.06 cm² |
| 26-mm Evolut R | 1.82 cm² | 1.69-2.10 cm² |
| 26-mm CoreValve | 2.01 cm² | 1.74-2.23 cm² |

**Table 1.** Simulated EOA comparisons to the acceptable ranges published in the literature [28,29] showing good agreement between the in vitro data and in silico results.

### *3.2 Hemodynamic Performance and Thrombogenic Risk of TAVR Leaflet Designs*

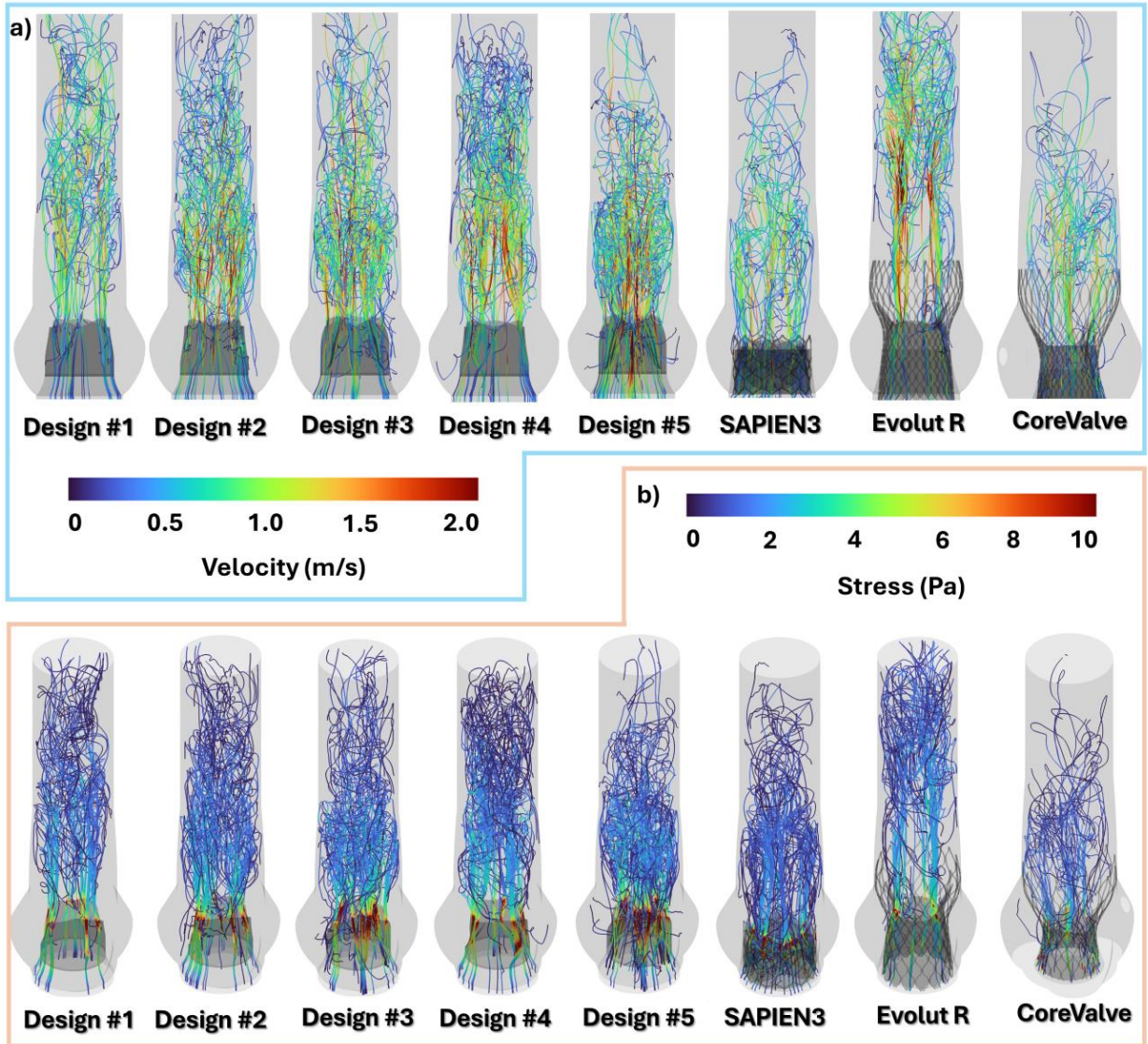

**Figure 9.** a) Trajectories for the top 100 platelets with highest stress accumulation – colored according to the instantaneous velocity along the trajectory); b) Additional depiction of the platelet trajectories – colored according to the corresponding stress the platelets are exposed to.

FSI videos show the hemodynamics and leaflet kinematics for each TAVR design and commercial device (Supplementary Video 1). Polymeric designs #1-2 and the SAPIEN3 exhibit lower peak fluid velocities compared to the other devices. This is likely due to their leaflets creating larger orifice areas and displaying reduced leaflet fluttering, which can be attributed to their more optimal surface area and leaflet flexion. For ease of viewing, we visualized the top 100 virtual platelets with highest cumulative stresses (Fig. 9). The trajectories (colored according to the instantaneous velocity along the trajectory) show peak values reaching around 2.5 m/s, with the highest platelet velocities reached near the leaflet tips and in the outflow regions just beyond the leaflets (Fig. 9a). We also overlay the stress attained by the virtual platelets at each instant along their path in the flow field (Fig. 9b). The platelet trajectories – colored according to the corresponding stress level that the platelets are exposed to, reach peak values around 10-12 Pa. The highest shear stresses on the platelets are seen at the leaflet surfaces, especially at the leaflet tips.

The conversion of the data into PDFs highlights varying thrombogenicity for differing valve designs (Fig. 10). Note these PDFs reflect platelet SA normalized to time due to the focus on one systolic pass of platelets through the various devices [16], allowing us to analyze the shear and principal stresses contributing to the total instantaneous stress loading on platelets due to differing valve designs. This precedes the full SA computation integrating the time component, which would skew results in this analysis due to the small time window of interest (systole). There are also no leak channels to analyze for platelet recirculation due to the idealized deployments. The commercial Evolut R, along with parametric designs #3-5, show the highest thrombogenic risk, given their longer PDF tail regions- indicating that more platelets are exposed to higher stress values. Studying the statistical distribution of the cumulative stresses the platelets are exposed to, the least and most thrombogenic devices and designs can be clearly identified (Table 2). The trileaflet design with minimized thrombogenic potential was design #1, while the most thrombogenic was design #5. For the commercial devices, the CoreValve was the least thrombogenic, while the Evolut R had the highest thrombogenic risk.

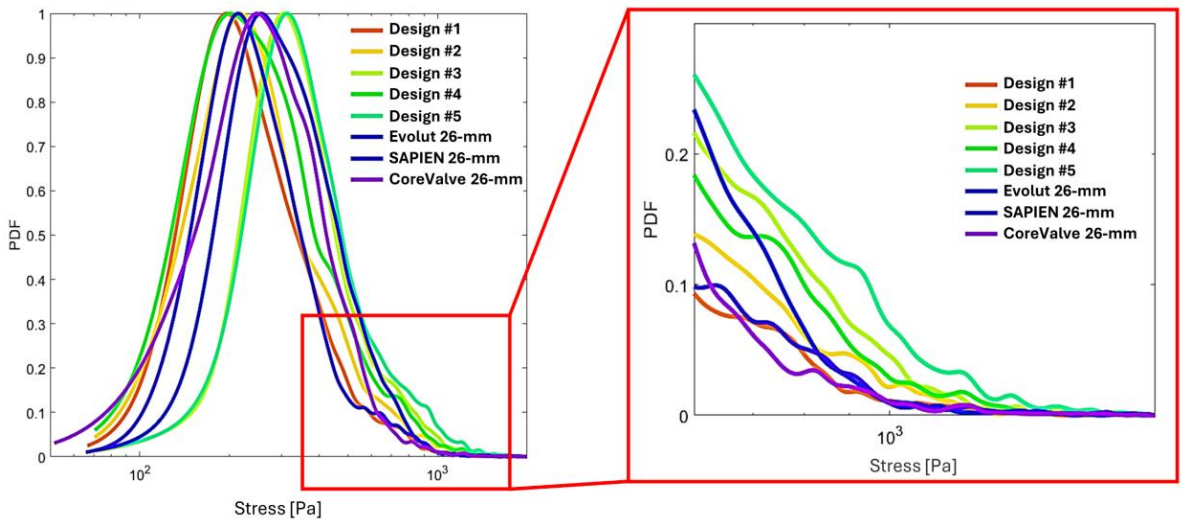

**Figure 10.** PDF of potential platelet activation for all five polymeric designs and three commercial devices. The inset figure zooms in on the tail region of the PDF curves to elucidate likelihood of platelets activating for each device. Note the PDFs presented here reflect the SA normalized to time, or the total stress amassed along each of the platelet trajectories for each simulation. This risk assessment differs slightly from the SA analysis conducted for the patient-specific models in the following section.

| Device | D1 | D2 | D3 | D4 | D5 | EV26 | SP26 | CV26 |
|---|---|---|---|---|---|---|---|---|
| Mean Stress Amassed Along Trajectories (Pa) | 319.5 | 350.0 | 435.8 | 367.7 | 468.0 | 382.9 | 337.2 | 342.9 |
| Median Stress Amassed Along Trajectories (Pa) | 264.9 | 286.1 | 371.8 | 302.7 | 390.7 | 343.3 | 274.0 | 305.4 |
| Probability of Finding Platelet in Tail Region (%) | 8.4 | 11.7 | 17.7 | 13.8 | 21.9 | 11.1 | 10.0 | 6.6 |
| | **Polymeric Valve Designs** | | | | | **Commercial Devices** | | |

**Table 2.** Stress statistics from top 100 "hot spot" platelet trajectories for each of the polymeric valve designs and commercial devices featuring bioprosthetic tissue leaflets. Abbreviations: D-Design, EV-Evolut R, SP-SAPIEN3, CV-CoreValve.

*3.3 TAVR-in-BAV: Clinical Parameters, Prosthetic Valve Performance and Thrombogenic Risk*

| Pt # | Deployment Eccentricity Index | GOA (cm$^2$) | EOA (cm$^2$) | Leak Volume (mL) | Stroke Volume (mL) | Cardiac Output (L/min) | Platelets Exceeding SA Threshold (%) | Median SA of Top 100 Hot Spot Trajectories (dyn-s/cm$^2$) |
|---|---|---|---|---|---|---|---|---|
| Pt. 1 | 1.94 | 1.92 | 1.27 | 7.17 | 53.48 | 2.69 | 0.65% | 70.75 |
| Pt. 2 | 1.05 | 2.07 | 1.35 | 2.49 | 55.12 | 3.06 | 0.005% | 11.87 |
| Pt. 3 | 1.21 | 1.96 | 1.37 | 4.93 | 55.70 | 2.97 | 0.56% | 78.39 |
| Pt. 4 | 1.16 | 2.01 | 1.43 | 0.231 | 59.98 | 3.52 | 0.03 | 19.52 |
| Pt. 5 | 1.67 | 1.51 | 0.95 | 0.201 | 37.86 | 2.21 | 0.005% | 11.55 |
| Pt. 6 | 1.11 | 2.11 | 1.51 | 4.91 | 65.66 | 3.56 | 0.6% | 68.78 |

**Table 3.** Simulation data for the six post-TAVR BAV patients. Deployment, device performance, relevant clinical parameters and thrombogenic risk are assessed.

Patient-specific hemodynamic and thrombogenic risk data were extracted from the simulations. The FSI videos demonstrating the post-TAVR patient simulations effectively showcase valve leaflets dynamics while also capturing leak flows during the cardiac cycle, which are crucial in assessing thrombogenic risk (Supplementary Video 2). Those provide valuable insights into the hemodynamic behavior of the prosthetic valves and their implications for patient outcomes, which are computed and compared quantitatively in detail (Table 3). Deployment eccentricity index was highest for patients #1, #3 and #5 having indices of 1.94, 1.21 and 1.67, respectively. Valve opening performance was assessed with geometric orifice area (GOA) and EOA. Patients #2 and #6 had the largest measured GOAs at 2.07 cm$^2$ and 2.11 cm$^2$, respectively (Fig. 11), while patients #4 and #6 had the largest measured EOAs at 1.43 cm$^2$ and 1.51 cm$^2$, respectively. The leak volume was 7.17, 2.49, 4.93, 0.231, 0.201 and 4.91 mL, respectively, for the patients in ascending order. Stroke volume and cardiac output were also evaluated, in addition to SA (Table 3). We again present the top 100 virtual platelets with highest SA (Fig. 12). The trajectories colored

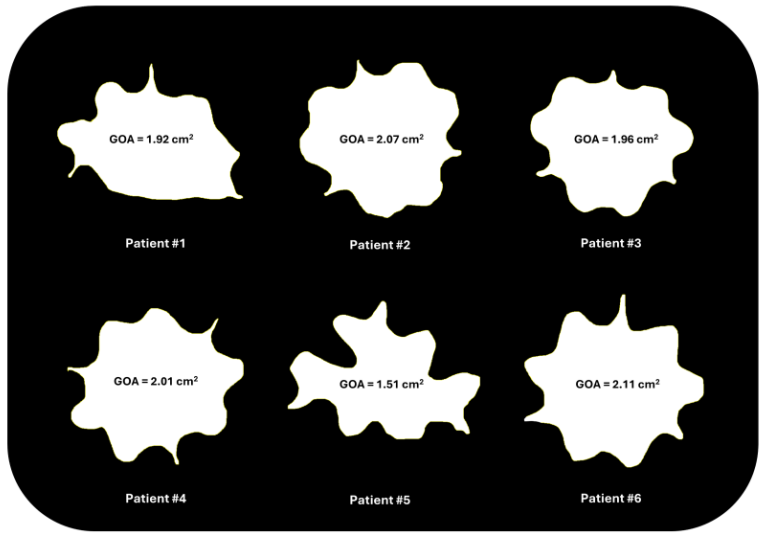

**Figure 11.** GOA of the 29-mm Evolut R valve post-TAVR in all six BAV patients, serving as a central parameter for assessing valve performance and efficiency.

by velocity show peak values reaching around 2.0 m/s (Fig. 12a). We also see the instantaneous stress value acting on the platelets (Fig. 12b), with peak values around 5-10 Pa, principally in the PVL gaps. The

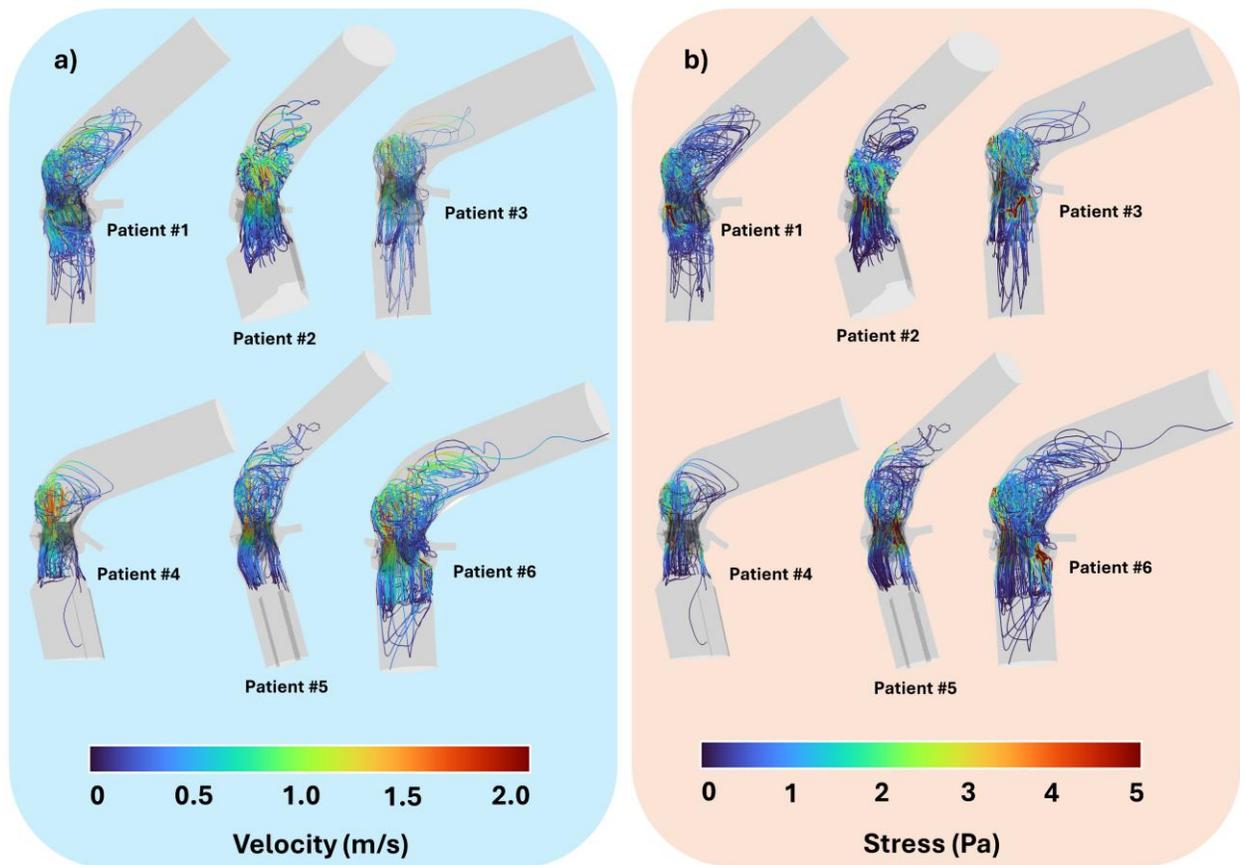

**Figure 12.** Thrombogenicity assessment of 29-mm Evolut R device in six BAV patient models. The top 100 platelet trajectories with overlaid contour plots of velocity and stress from each simulation are shown.

PDFs computed for the BAV patients ("thrombogenic footprint") show varying levels of thrombogenic risk, where we see "spiking" behavior in the tail region for all the patient cases (Fig. 13). The thrombogenic risk analysis shows that patients #1, #3 and #6 are at highest risk of thrombosis due to activating potential of platelets in the flow, with the tail regions for these simulated patients (Fig. 13, inset zoom in) show the largest degree of PDF spiking. The tabulated data correlates well with this, where the percentage of platelets exceeding the Hellums activation threshold (35 dyn-s/cm$^2$) was 0.65%, 0.56% and 0.6% for patients #1, #3 and #6, respectively. The other patients had percentages of 0.03% or lower. The median SA values for the top 100 "hot spot" trajectories for each patient also show the same trend, where patients #1, #3 and #6 had median SA values of 70.75, 78.39 and 68.78 dyn-s/cm$^2$, respectively. The other patients had lower median values but still clearly showed platelets exceeding the SA threshold, highlighting the varying extent of thrombosis risk between patients.

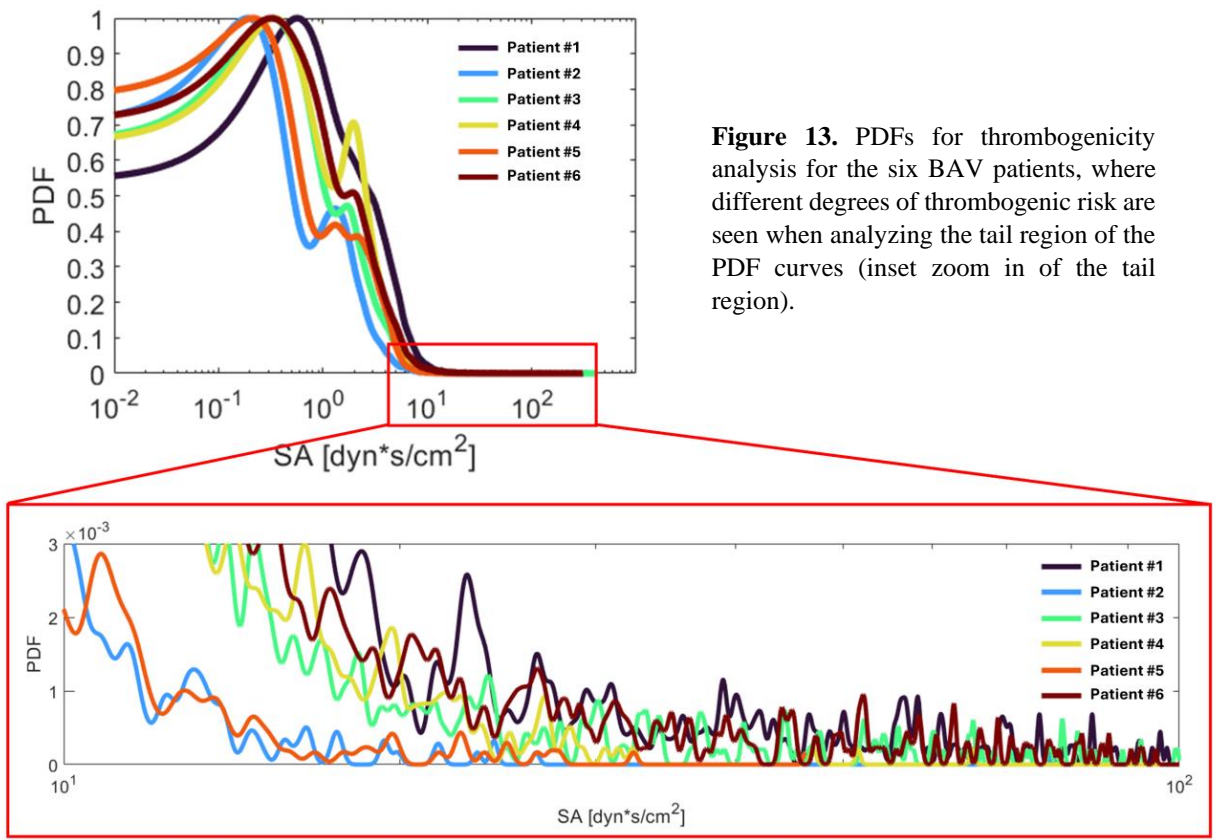

**Figure 13.** PDFs for thrombogenicity analysis for the six BAV patients, where different degrees of thrombogenic risk are seen when analyzing the tail region of the PDF curves (inset zoom in of the tail region).

## 4. DISCUSSION

The study outlined here consisted of three key phases. First, an FSI workflow was created and validated with in vitro data. EnSight was utilized for virtual platelet seeding and a thrombogenicity analysis method that has been previously validated [15-17]. Subsequently, this workflow was employed in an engineering R&D approach to investigate the impact of different TAVR leaflet designs on valve performance and thrombogenic risk. Lastly, the simulation workflow was utilized to examine the clinical perspective of TAVR valve performance and thrombogenic risk in six patient-specific models. Employing pulsatile flow loop experiments, conducted by us and others, is the standard regulated approach of testing prosthetic valves and validating the simulation predictions [16, 21, 35]. Some minor differences between the in silico model and the benchtop experiment include: i) in silico flow rate slightly overshooting the peak experimental flow rate curve, and ii) observed oscillations in the simulated flow rate during diastole, which has been observed previously [23, 36] (Figures 6, 7). Despite these differences, the primary flow properties governing valve dynamics are accurately captured.

Previous TAVR valve optimization studies have focused on stress distribution in the leaflets [37, 38]. In our leaflet design study, our focus was on the fact that the majority of current TAVR devices are stented, bioprosthetic tissue valves, which have limitations in durability and design. Novel polymeric TAVR valves offer a promising alternative, allowing for engineering flexibility, hemocompatibility, structural integrity and reproducibility [21]. Optimization can be achieved via geometric modifications to the polymeric designs. It offers design freedom that is very limited in harvested tissue specimens used to fabricate bioprosthetic TAVR devices. By employing our FSI approach, we further demonstrate how thrombogenicity can be used as an additional design optimization parameter. This is especially important when the rapid expansion of TAVR into younger and lower-risk patients is considered. Our simulation results depict highest stresses forming near the surfaces of the prosthetic leaflets and at the leaflet tips. This demonstrates the key idea that platelets interact differently with varying valve designs and experience different magnitudes of stress along their flow trajectories past each prosthetic device design. Polymeric design #1 and #2, in addition to the 26-mm SAPIEN3 and CoreValve devices, showed the lowest peak flow velocities when compared to the other devices, attributed to these valve designs producing larger orifice areas and having reduced leaflet fluttering [39, 40], likely due to more optimal surface area and leaflet flexion. Regarding thrombogenic risk, polymeric design #1 and the 26-mm CoreValve had the lowest thrombogenic footprints for the polymeric valve designs and commercial devices, respectively, likely attributed to more optimal EOA, leaflet flexion and folding, which vary based on leaflet geometry. This variation impacts how unobstructed the platelet trajectories are and their cumulative stresses. Appropriately optimized, novel polymeric devices are likely to outperform current tissue devices and displace them.

Previous in silico models have utilized CFD approaches or other simplified models which fail to capture heart valve dynamics, mostly focusing on flow stasis as thrombus promoter- neglecting flow induced stresses acting on the flowing platelets, or rely on benchtop experiments with inherent system limitations and noisy data [9, 18, 19, 41]. Additionally, we previously analyzed thrombogenicity using weak FSI coupling techniques limited in analysis duration, and without consideration for TAVR devices [16]. We have now advanced to our current strong coupled FSI-thrombogenicity approach which overcomes these limitations, as demonstrated by its effectiveness in studying post-TAVR outcomes in patient-specific models as well. Specifically, we examined TAVR device outcomes in patients with BAV disease since TAVR devices that were designed for the circular trileaflet aortic valve anatomies often deploy elliptically in these patients, and as a result experience worse TAVR outcomes, including increased risk of thrombosis [22, 42]. The methodology developed in our study can uncover key clinical parameters that may be critical for both assessing and mitigating thrombogenic risk, along with other essential hemodynamic criteria. It can also establish correlations between numerous parameters such as device EOA and patient stroke volume or cardiac output. When we observed a reduction in the orifice area by nearly 30-40%, it resulted in a corresponding decrease in these clinical parameters by approximately the same percentage. Our study underscores a significant association between increased leak flow and heightened thrombogenic risk, aligning with previous investigations [18, 19]. Despite the leak flows observed here being clinically categorized as "mild", notable discrepancies persisted regarding its associated thrombogenic risk. This underestimation and discrepancy between the clinical category and the thrombogenic potential was observed clinically in TAVR patients [19]. These findings suggest a nuanced interplay of patient-specific factors and device characteristics that warrant further investigation for comprehensive risk management strategies. The quantitative information provided by our analysis can inform procedural planning, allowing clinicians to tailor interventions more effectively, based on patient-specific characteristics.

Our study had several limitations. Idealized TAVR deployments were generated with no leak channels in the first part of this study, but this was strategically done for design comparison purposes. Due to the computational expense of simulations, two cardiac cycles were simulated for each in silico experiment. More cardiac cycles are needed to better assess recirculating platelets. In this study, whether the simulated anatomies were idealized or patient-specific, rigid walls were used in the fluid domain setup, as their compliance was assumed negligible during the cardiac cycle. Incorporating elastic walls into our FSI simulations may lead to more accurate predictions of wall shear stress, pressure distribution, and aortic flow profiles, but does not hinder the cardiac parameters highlighted in this work. For the patient-specific simulations, we only focused on type I BAV patients and one TAVR device type. This will be expanded and diversified in future studies to increase the applicability of our approach. Also, while pre-TAVR CT scans and device deployment data were available to replicate the patient-specific TAVR deployments virtually, patient-specific pressure conditions from cardiac catheterization were not available to use as boundary conditions in the FSI studies. Given these factors, we successfully achieved the primary objective of using our FSI-thrombogenicity methodology to assess the relative thrombogenic differences between TAVR device designs and patient-specific scenarios.

## 5. CONCLUSION

Our integrated FSI approach aids in the assessment and minimization of prosthetic valve thrombogenicity from the initial design stage to predicting its potential application in the clinic. It allows for unparalleled optimization of polymeric TAVR devices, uncovering key leaflet design parameters that may be critical for mitigating thrombogenic risk, in addition to patient-specific modeling to evaluate device performance once implanted. This may be expanded to assess thrombogenic risk of numerous cardiovascular implants. Optimized TAVR devices of the future may finally reduce the need for post-TAVR antithrombotic therapies and redo TAVR procedures once the devices fail. Future work will include analyzing flow stasis and residence time of the virtual platelets to address other poor clinical outcomes related to TAVR, such as hypoattenuated leaflet thickening (HALT). Our in silico approach facilitates conducting extensive testing without the need for costly and time-consuming in vitro experiments. This complementary approach to the current in vitro testing may eventually be established as an accurate, reliable, and cost-effective alternative that will displace it. This supports the future era of clearance and approval by regulatory bodies such as the Food and Drug Administration (FDA), using in silico testing for medical devices and "virtual twin" approaches to clinical trials and medicine.

## 6. ACKNOWLEDGEMENTS

Funding: This work was supported by the National Institutes of Health grants: NIBIB: BRP U01 EB026414 (DB), and NHLBI: R41 (DB) HL134418 and R42 HL134418 (DB).

Industry Partners: ANSYS, SIMULIA

Disclosure: Author KB is a consultant for PolyNova Cardiovascular Inc. Authors DB and MS have an equity interest in PolyNova Cardiovascular Inc.

## 7. REFERENCES

1.	Parikh, P.B., et al., *Transcatheter aortic valve replacement in heart failure.* European Journal of Heart Failure, 2024.
2.	Raghav, V., et al., *Transcatheter aortic valve thrombosis: a review of potential mechanisms.* Journal of the Royal Society Interface, 2021. **18**(184): p. 20210599.
3.	Fukumoto, Y., *Thrombogenicity After Transcatheter Aortic Valve Replacement.* American Journal of Cardiology, 2023. **207**: p. 257-259.
4.	Avvedimento, M., et al., *Bleeding events after transcatheter aortic valve replacement: JACC state-of-the-art review.* Journal of the American College of Cardiology, 2023. **81**(7): p. 684-702.
5.	Maznyczka, A. and T. Pilgrim, *Antithrombotic Treatment After Transcatheter Valve Interventions: Current Status and Future Directions.* Clinical therapeutics, 2023.
6.	Llerena-Velastegui, J., et al., *Frequency of Stroke in Intermediate-Risk Patients in the Long Term Undergoing TAVR vs. SAVR: A Systematic Review and Meta-Analysis.* Current Problems in Cardiology, 2023: p. 102099.
7.	Di Pietro, G., et al., *Stroke Risk during TAVR: Is Prevention Better than Cure?*, in *Aortic Valve Disease-Recent Advances*. 2023, IntechOpen.
8.	Pislaru, S.V., V.T. Nkomo, and G.S. Sandhu, *Assessment of prosthetic valve function after TAVR.* JACC: Cardiovascular Imaging, 2016. **9**(2): p. 193-206.
9.	Vahidkhah, K., et al., *Valve thrombosis following transcatheter aortic valve replacement: significance of blood stasis on the leaflets.* European Journal of Cardio-Thoracic Surgery, 2017. **51**(5): p. 927-935.
10.	*ISO 5840-3 (2021), ICS:11.040.40*.
11.	Braune, S., et al., *In vitro thrombogenicity testing of biomaterials.* Advanced healthcare materials, 2019. **8**(21): p. 1900527.
12.	Corrigan, F.E., et al., *Imaging for predicting, detecting, and managing complications after transcatheter aortic valve replacement.* JACC: Cardiovascular Imaging, 2019. **12**(5): p. 904-920.
13.	*Credibility of Computational Models Program: Research on Computational Models and Simulation Associated with Medical Devices*. 11/16/2023; Available from: https://www.fda.gov/medical-devices/medical-device-regulatory-science-research-programs-conducted-osel/credibility-computational-models-program-research-computational-models-and-simulation-associated.
14.	Bluestein, D., S. Einav, and M.J. Slepian, *Device thrombogenicity emulation: a novel methodology for optimizing the thromboresistance of cardiovascular devices.* Journal of biomechanics, 2013. **46**(2): p. 338-344.
15.	Xenos, M., et al., *Device Thrombogenicity Emulator (DTE)− Design optimization methodology for cardiovascular devices: A study in two bileaflet MHV designs.* Journal of biomechanics, 2010. **43**(12): p. 2400-2409.
16.	Piatti, F., et al., *Hemodynamic and thrombogenic analysis of a trileaflet polymeric valve using a fluid–structure interaction approach.* Journal of biomechanics, 2015. **48**(13): p. 3641-3649.
17.	Chiu, W.C., et al., *Device thrombogenicity emulation: an in silico predictor of in vitro and in vivo ventricular assist device thrombogenicity.* Scientific Reports, 2019. **9**(1): p. 2946.
18.	Anam, S.B., et al., *Assessment of paravalvular leak severity and thrombogenic potential in transcatheter bicuspid aortic valve replacements using patient-specific computational modeling.* Journal of cardiovascular translational research, 2021: p. 1-11.
19.	Kovarovic, B.J., et al., *Mild paravalvular leak may pose an increased thrombogenic risk in transcatheter aortic valve replacement (TAVR) patients-insights from patient specific in vitro and in silico studies.* Bioengineering, 2023. **10**(2): p. 188.

20. Spears, J., et al., *TAVR: a review of current practices and considerations in low-risk patients.* Journal of interventional cardiology, 2020. **2020**.
21. Kovarovic, B., et al., *Visions of TAVR future: Development and optimization of a second generation novel polymeric TAVR.* Journal of Biomechanical Engineering, 2022. **144**(6): p. 061008.
22. Al-Asad, K.S., et al., *Transcatheter Aortic Valve Replacement in Bicuspid Versus Tricuspid Aortic Valve Stenosis: Meta-Analysis and Systemic Review.* The American Journal of Cardiology, 2023. **203**: p. 105-112.
23. Oliveira, H.L., et al., *Three-dimensional fluid–structure interaction simulation of the Wheatley aortic valve.* International Journal for Numerical Methods in Biomedical Engineering, 2024. **40**(2): p. e3792.
24. Liu, X., et al., *Fluid-structure interaction analysis on the influence of the aortic valve stent leaflet structure in hemodynamics.* Frontiers in physiology, 2022. **13**: p. 904453.
25. Hoeijmakers, M.J.M.M., *Analysis of aortic-valve blood flow using computational fluid dynamics.* 2021.
26. Zhou, J., et al., *Analysis of the Effect of Thickness on the Performance of Polymeric Heart Valves.* Journal of Functional Biomaterials, 2023. **14**(6): p. 309.
27. Li, K. and W. Sun, *Simulated thin pericardial bioprosthetic valve leaflet deformation under static pressure-only loading conditions: implications for percutaneous valves.* Annals of biomedical engineering, 2010. **38**: p. 2690-2701.
28. Hatoum, H., et al., *The hemodynamics of transcatheter aortic valves in transcatheter aortic valves.* The Journal of thoracic and cardiovascular surgery, 2021. **161**(2): p. 565-576. e2.
29. Hahn, R.T., et al., *Comprehensive echocardiographic assessment of normal transcatheter valve function.* JACC: Cardiovascular Imaging, 2019. **12**(1): p. 25-34.
30. Apel, J., et al., *Assessment of hemolysis related quantities in a microaxial blood pump by computational fluid dynamics.* Artificial organs, 2001. **25**(5): p. 341-347.
31. Dumont, K., et al., *Comparison of the hemodynamic and thrombogenic performance of two bileaflet mechanical heart valves using a CFD/FSI model.* 2007.
32. Helbock, R.T., et al., *Designing a novel asymmetric transcatheter aortic valve for stenotic bicuspid aortic valves using patient-specific computational modeling.* Annals of Biomedical Engineering, 2023. **51**(1): p. 58-70.
33. Gregg, D.E., *Phasic blood flow and its determinants in the right coronary artery.* American Journal of Physiology-Legacy Content, 1937. **119**(3): p. 580-588.
34. Barash, P.G., *Clinical anesthesia*. 2009: Lippincott Williams & Wilkins.
35. Stasiak, J.R., et al., *Design, development, testing at ISO standards and in vivo feasibility study of a novel polymeric heart valve prosthesis.* Biomaterials Science, 2020. **8**(16): p. 4467-4480.
36. Hsu, M.-C., et al., *Fluid–structure interaction analysis of bioprosthetic heart valves: significance of arterial wall deformation.* Computational mechanics, 2014. **54**: p. 1055-1071.
37. Travaglino, S., et al., *Computational optimization study of transcatheter aortic valve leaflet design using porcine and bovine leaflets.* Journal of biomechanical engineering, 2020. **142**(1): p. 011007.
38. Abbasi, M. and A.N. Azadani, *A geometry optimization framework for transcatheter heart valve leaflet design.* Journal of the mechanical behavior of biomedical materials, 2020. **102**: p. 103491.
39. Johnson, E.L., et al., *Thinner biological tissues induce leaflet flutter in aortic heart valve replacements.* Proceedings of the National Academy of Sciences, 2020. **117**(32): p. 19007-19016.
40. Iseki, Y., et al., *Fluttering Bioprosthetic Valve Leaflet Detected by Intravascular Ultrasound During Valve-in-Valve Transcatheter Aortic Valve Replacement.* Case Reports, 2021. **3**(6): p. 910-912.

41. Borowski, F., et al., *Analysis of thrombosis risk of commissural misaligned transcatheter aortic valve prostheses using particle image velocimetry.* Tm-technisches Messen, 2023(0).
42. Improta, R., et al., *A Meta-Analysis of Short-Term Outcomes of TAVR versus SAVR in Bicuspid Aortic Valve Stenosis and TAVR Results in Different Bicuspid Valve Anatomies.* Journal of Clinical Medicine, 2023. **12**(23): p. 7371.